\documentclass[11pt, letterpaper]{amsart}
\usepackage{graphicx, amssymb}
\usepackage{amsmath}

\newtheorem{thm}{Theorem}[section]

\newtheorem{lem}[thm]{Lemma}
\newtheorem{prop}[thm]{Proposition}
\theoremstyle{definition}

\theoremstyle{remark}
\newtheorem{rem}[thm]{Remark}
\newtheorem{exa}[thm]{Example}
\numberwithin{equation}{section}

\newcommand{\set}[1]{\left\{#1\right\}}
\newcommand{\Real}{\mathbb R}
\newcommand{\Natural}{\mathbb N}

\newcommand{\such}{\, | \, }
\newcommand{\prob}{\mathbb{P}}

\newcommand{\Time}{\mathfrak{T}}

\newcommand{\qprob}{\mathbb{Q}}
\newcommand{\expec}{\mathbb{E}}

\newcommand{\expecq}{\expec_\qprob}

\newcommand{\basis}{(\Omega,  \, (\F_t)_{t \in \Real_+}, \, \prob)}

\newcommand{\F}{\mathcal{F}}

\newcommand{\cadlag}{c\`adl\`ag}

\newcommand{\ud}{\mathrm d}

\newcommand{\liminfn}{\liminf_{n \to \infty}}
\newcommand{\limsupn}{\limsup_{n \to \infty}}

\newcommand{\limn}{\lim_{n \to \infty}}

\newcommand{\ttau}{{\widetilde{\tau}}}

\newcommand{\uS}{\underline{S}}
\newcommand{\oS}{\overline{S}}

\newcommand{\uC}{\underline{\mathcal{C}}}
\newcommand{\oC}{\overline{\mathcal{C}}}

\newcommand{\pare}[1]{\left(#1\right)}
\newcommand{\bra}[1]{\left[#1\right]}
\newcommand{\dbra}[1]{[\kern-0.15em[ #1 ]\kern-0.15em]}
\newcommand{\dbraco}[1]{[\kern-0.15em[ #1 [\kern-0.15em[}
\newcommand{\dbraoc}[1]{]\kern-0.15em] #1 ]\kern-0.15em]}

\newcommand{\og}{\overline{g}}

\newcommand{\dfn}{\, := \,}

\newcommand{\htau}{\widehat{\tau}}

\newcommand{\indic}{\mathbb{I}}

\newcommand{\tsigma}{\widetilde{\sigma}}

\newcommand{\esssup}[1]{{\text{esssup}_{#1}}}

\title[]{Strict Local Martingale Deflators and Pricing American Call-Type Options}\author[]{Erhan Bayraktar} \thanks{
We also thank the seminar participants at Cornell University, Georgia State University, Rutgers University, the University of California at Santa Barbara, and the University of Michigan.}
\thanks{E. Bayraktar is supported in part by the National Science Foundation under grant number DMS-0906257.}
\address[Erhan Bayraktar]{Department of Mathematics, University of Michigan, 530 Church Street, Ann Arbor, MI 48104, USA}
\email{erhan@umich.edu}
\author[]{Constantinos Kardaras}\thanks{C. Kardaras is supported in part by the National Science Foundation under grant number DMS-0908461}
\address[Constantinos Kardaras]{Department of Mathematics and Statistics, Boston University, 111 Cummington Street, Boston, MA 02215, USA}
\email{kardaras@bu.edu}
\author[]{Hao Xing }
\address[Hao Xing]{Department of Mathematics and Statistics, Boston University, 111 Cummington Street, Boston, MA 02215, USA}
\email{haoxing@bu.edu}

\date{December 21, 2009}
\begin{document}

\begin{abstract}
We solve the problem of pricing and optimal exercise of American call-type options in markets which do not necessarily admit an equivalent local martingale measure. This resolves an open question proposed by Fernholz and Karatzas [Stochastic Portfolio Theory: A Survey, \emph{Handbook of Numerical
Analysis}, 15:89Ð168, 2009]. \\ \\
\textbf{Key Words:} Strict local martingales, deflators, American call options.

\end{abstract}
\maketitle

\setcounter{section}{-1}

\section{Introduction}

Let $\beta$ be a strictly positive and nonincreasing process with $\beta_0 = 1$ and $S$ be a strictly positive semimartingale. In financial settings, $S$ is the price of an underlying asset and $\beta$ the discount factor, i.e., the reciprocal of the value of the savings account. We shall assume throughout that there exists a strictly positive local martingale $Z$ with $Z_0=1$ such that $L \dfn Z \beta S$ is a local martingale. For simplicity in notation, we also denote $Y \dfn Z \beta$, which is a supermartingale. Let $g: \Real_+ \mapsto \Real_+$ be a nonnegative \emph{convex} function with $g(0) = 0$, $g(x) < x$ holding for some (and then for all) $x \in \Real_{++} \equiv \Real_+ \setminus \set{0}$, and $\lim_{x \uparrow \infty} \pare{g(x) / x} = 1$. The canonical example of such function is $g(x) = (x - K)_+$ for $x \in \Real_+$, where $K \in \Real_{++}$ --- this will correspond to the payoff function of an American call option in the discussion that follows. With $X \dfn Y g(S)$, we consider the following optimization problem:
\begin{equation} \label{eq: problem} \tag{OS}
\text{Compute } v \dfn \sup_{\tau \in \Time}\expec\left[X_{\tau}\right], \text{ and find } \htau \in \Time \text{ such that } \expec[X_{\htau}] = v,
\end{equation}
where $\Time$ is used to denote the set of all stopping times (we consider also infinite-valued ones; we shall see later that $X_\infty \dfn \lim_{t \to \infty} X_t$ is well-defined and $\prob$-a.s. finite). A stopping time $\htau \in \Time$ such that $\expec\left[X_{\htau}\right] = v$ will be called \textsl{optimal} for the problem \eqref{eq: problem}. Since $g(x) \leq x$ for all $x \in \Real_+$, $X \leq Y S =L$. As a result, $v \leq L_0 = S_0 < \infty$.

Although the problem is defined for infinite time-horizon, it includes as a special case any finite-time-horizon formulation. If the time-horizon is $T \in \Time$, then just consider the processes $Z$, $S$, and $\beta$ as being constant for all times after $T$ on the event $\{T< \infty\}$.

\smallskip

Only for the purposes of this introductory discussion, and in order to motivate the study of the problem \eqref{eq: problem}, we  assume that $Z$ is the unique local martingale that makes $Z \beta S$ a local martingale, as well as that the time-horizon of the problem is $T \in \Time$, where $\prob [T < \infty] = 1$. Here, $T$ denotes the \emph{maturity} of an American claim with immediate payoff $g(S_\tau)$ when it is exercised at time $\tau \in \Time$ with $\tau \leq T$. It follows from Corollary~2.1 of \cite{Stricker-Yan} that  any contingent claim with maturity $T$ can be perfectly hedged by dynamically trading in $S$ and in the savings account. As a result, the market is still \textit{complete}, even though the local martingale $Z$ may be a strict local martingale. (See also Section 10 of \cite{Fernholz-Karatzas-survey}.) There exists a \cadlag \ process $A = (A_t)_{t \in [0, T]}$ such that $A_t = Y_t^{-1} \esssup{\tau \in \Time{[t, T]}} \expec\left[X_{\tau}\, \big| \, \F_{t}\right]$ for $t<T$ and $A_T =  g\left(S_T\right)$ (see Lemma~2.4 in \cite{Stricker-Yan}). Here, for any $s \in \Time$ and $\tau \in \Time$, we let $\Time{[s, \tau]}$ denote the class of all $\tau' \in \Time$ with $s \leq \tau' \leq \tau$. Remark~2 in \cite{Stricker-Yan} shows that $A$ is the smallest process with the following two properties: (a) $A$ dominates $g(S)$ over $[0,T]$, and (b) $Z \beta A$ is supermartingale. Additionally, thanks to the \textit{optional decomposition theorem} (see Theorem~2.1 of \cite{Stricker-Yan} and the references therein for the details), there exist a predictable process $\phi$ and an adapted nonnegative and nondecreasing process $C$ with $C_0=0$, such that $\beta A = A_0 + \int_0^\cdot \phi_t \ud (\beta_t S_t) - C$. As a result, $v=A_0$ is the upper hedging price of the American option with the associated payoff function $g$.

When there is a stock price \emph{bubble}, i.e., $Z \beta S$ is a strict local martingale, or
when the market allows for \emph{arbitrage} opportunities relative to the bank account, i.e., $Z$ is a strict local martingale, the prices of derivative securities present several anomalies because of the lack of martingale property
--- see, for example, \cite{Cox-Hobson}, \cite{Madan-Yor},\cite{Heston-Loewenstein-Willard}, \cite{MR2359365}, \cite{jps}, \cite{Pal-Protter}, \cite{DFern-Kar09}, and \cite{Fernholz-Karatzas-survey}.
One instance of such an anomaly is the failure of the \emph{put-call parity}. Our goal in this paper is to examine in detail the optimization problem \eqref{eq: problem}, and, therefore, gain a better understanding of pricing of American options of the ``call'' type in markets as discussed above. This resolves an open question put forth in \cite{Fernholz-Karatzas-survey}, Remark 10.4. We give a condition that, when satisfied, guarantees the existence of optimal stopping times for \eqref{eq: problem}. If this condition is satisfied, we give an explicit expression for the smallest optimal stopping time. On the other hand, when this condition is violated, there are no optimal stopping times.
Although there is generally no optimal exercise time, there are examples in which one prefers to exercise early. This should be contrasted to Merton's \emph{no early exercise theorem} which applies when both $Z$ and $Z \beta S$ are martingales. (This theorem states that it is optimal to exercise an American call option only at maturity.) The expression for $v$ given in Theorem~\ref{thm: main}, our main result, generalizes Theorem A.1 of \cite{Cox-Hobson} and Proposition 4.1 of \cite{Heston-Loewenstein-Willard} to a setting where the underlying process is a general semimartingale with possibly discontinuous paths and the market does not necessarily admit a local martingale measure.

The structure of the paper as follows. In Section~\ref{sec:main}, we give our main result, whose proof is deferred to Section~\ref{sec: proof}. In Section~\ref{sec:examples}, we give examples to illustrate our findings.

\section{The Main Result}\label{sec:main}

\subsection{The set-up}

Throughout, we shall be working on a filtered probability space $\basis$. The filtration $(\F_t)_{t \in \Real_+}$ is assumed to be right-continuous and $\prob$ is a probability on $(\Omega, \F_\infty)$, where $\F_{\infty}:=\bigvee_{t \in \Real_+}\F_t$. All relationships between random variables will be understood in the $\prob$-a.s. sense. All processes that appear in the sequel are assumed \cadlag \ unless noted otherwise; (in)equalities involving processes are supposed to hold everywhere with the possible exception of an evanescent set.

We keep all the notation from the Introduction. In particular, $\Time$ denotes the set of \emph{all} (possibly infinite-valued) stopping times with respect to $(\F_t)_{t \in \Real_+}$. In order to make sense of the problem described in \eqref{eq: problem}, we must ensure that $X_\infty \dfn \lim_{t \to \infty} X_t$ exists, which we shall do now. As both $Z$ and $L$ are nonnegative local martingales, and in particular supermartingales, the random variables $Z_{\infty} \dfn \lim_{t\rightarrow \infty} Z_t$ and $L_\infty \dfn \lim_{t \to \infty} L_t$ are well-defined and finite. In a similar manner we also define $\beta_\infty \dfn \lim_{t\rightarrow \infty} \beta_t$. Define now a new function $h : \Real_+ \cup \set{\infty} \mapsto \Real_+$ via $h(x) \dfn g(x) / x$ for $x \in \Real_{++}$, and $h(0) = \lim_{x \downarrow 0} h(x)$ as well as $h(\infty) = \lim_{x \uparrow \infty} h(x)$. Given the properties of $g$, it is plain to see that $h$ is nonnegative, nondecreasing, $0 \leq h \leq 1$, and $h(\infty) = 1$. Furthermore, $X = L \, h(S)$. Now, on $\set{L_\infty > 0}$ we have  $ \lim_{t \to \infty} S_t = L_\infty /  Y_\infty$ which takes values in $\Real_{++} \cup \set{\infty}$; therefore, on $\set{L_\infty > 0}$ we have $X_\infty = L_\infty h (S_\infty)$. On the other hand, it is clear that on $\set{L_\infty = 0}$ we have $X_\infty = 0$, since $h$ is a bounded function.

\subsection{The default function}

Recall that a nondecreasing sequence of stopping times $(\sigma^n)_{n\in \Natural}$ is a \emph{localizing sequence} for $L$ if $(L_{\sigma^n \wedge t })_{t \in \Real_+}$ is a \emph{uniformly integrable} martingale for all $n \in \Natural$, and $\uparrow \limn \sigma^n = \infty$. The following result will allow us to define a very important function.

\begin{lem} \label{lemma:delta}
Let $\tau \in \Time$ and consider any localizing sequence $(\sigma^n)_{n\in \Natural}$ for $L$. Then, the sequence $\left( \expec[X_{\tau \wedge \sigma^n}] \right)_{n \in \Natural}$ is nondecreasing. Furthermore, the quantity
\begin{equation}\label{eq: def delta}
 \delta(\tau) \dfn \uparrow \lim_{n\rightarrow \infty} \expec[X_{\tau \wedge \sigma^n}] - \expec[X_\tau] = \lim_{n\rightarrow \infty} \expec\left[X_{\sigma^n} \indic_{\{\tau>\sigma^n\}} \right]
\end{equation}
is nonnegative and independent of the choice of the localizing sequence $(\sigma^n)_{n\in \Natural}$ for $L$.
\end{lem}

\begin{proof}
Let $\og :\Real_+ \mapsto \Real_+$ be defined via $\og(x) = x - g(x)$ for $x \in \Real_+$; $\og$ is nonnegative, nondecreasing, and concave. Define also $W \dfn Y \og(S)$, so that $X = L - W$. First, let us show that $W$ is a nonnegative supermartingale. Let $(s^n)_{n \in \Natural}$ be a localizing sequence for $Z$. For each $n \in \Natural$, we define a probability $\mathbb{Q}^n \sim \prob$ on $(\Omega, \F_{\infty})$ via $\ud \mathbb{Q}^n = Z_{s^n} \ud \prob$. Let $u \in \Real_+$ and $t \in \Real_+$ with $u \leq t$. Then
\[
\begin{split}
&\expec\left[Z_t \frac{\beta_t}{\beta_u} \og(S_t)\bigg|\F_u\right] = \expec\left[\lim_{n \to \infty}Z_{t \wedge s^n} \frac{\beta_{t \wedge s^n}}{\beta_{u \wedge s^n}} \og(S_{t \wedge s^n})\bigg|\F_u\right] \leq \liminf_{n \to \infty}  \expec\left[Z_{t \wedge s^n} \frac{\beta_{t \wedge s^n}}{\beta_{u \wedge s^n}} \og(S_{t \wedge s^n})\bigg|\F_u\right] \\
&=  \liminf_{n \to \infty}  Z_{u \wedge s^n} \mathbb{E}_{\mathbb{Q}^n}\left[ \frac{\beta_{t \wedge s^n}}{\beta_{u \wedge s^n}} \og(S_{t \wedge s^n})\bigg|\F_{u \wedge s^n} \right]  \leq  \liminf_{n \to \infty} Z_{u \wedge s^n} \mathbb{E}_{\mathbb{Q}^n}\left[ \og \left( \frac{\beta_{t \wedge s^n} S_{t \wedge s^n}}{\beta_{u \wedge s^n}} \right) \bigg|\F_{u \wedge s^n} \right]
\\&\leq \liminf_{n \to \infty}   Z_{u \wedge s^n} \og \left(\mathbb{E}_{\mathbb{Q}^n}\left[\frac{\beta_{t \wedge s^n} S_{t \wedge s^n}}{\beta_{u \wedge s^n}} \bigg|\F_{u \wedge s^n} \right]\right) \leq \lim_{n \to \infty} Z_{u \wedge s^n} \og \left(S_{u \wedge s^n}\right) =Z_u \og(S_u).
\end{split}
\]
Here, the first inequality follows from Fatou's lemma. The second inequality follows from the concavity of $\og$, noting that $\og(0) = 0$. The third inequality, on the other hand, is thanks to Jensen's inequality. The last inequality is due to the fact that $\og$ is non-decreasing and $\mathbb{E}_{\mathbb{Q}^n}\left[\beta_{t \wedge s^n} S_{t \wedge s^n} \such \F_{u \wedge s^n} \right]\leq \beta_{u \wedge s^n} S_{u \wedge s^n}$.

Let $(\sigma^n)_{n \in \Natural}$ be any localizing sequence for $L$. To see that $(\expec[X_{\tau \wedge \sigma^n}])_{n \in \Natural}$ is nondecreasing, simply write $\expec[X_{\tau \wedge \sigma^n}] = \expec\left[L_{\tau \wedge \sigma^n} - W_{\tau \wedge \sigma^n} \right] = L_0 - \expec\left[W_{\tau \wedge \sigma^n} \right]$ and use the supermartingale property of $W$.

Now, pick another localizing sequence $(\tilde{\sigma}^m)_{m\in \Natural}$ for $L$. For fixed $n \in \Natural$, $\set{Y_{\tsigma \wedge \sigma^n} S_{\tsigma \wedge \sigma^n} \such \tsigma \in \Time}$ is a uniformly integrable family. Since $0 \leq  X_{\tsigma \wedge \sigma^n} \leq Y_{\tsigma \wedge \sigma^n} S_{\tsigma \wedge \sigma^n}$ holds for all $n \in \Natural$ and $\tsigma \in \Time$, $\set{ X_{\tsigma \wedge \sigma^n}   \such \tsigma \in \Time}$ is also a uniformly integrable family for all $n \in \Natural$. Similarly, we can also derive that $\set{ X_{\sigma \wedge \tilde{\sigma}^m}  \such \sigma \in \Time}$ is a uniformly integrable family for all $m \in \Natural$. It follows that both processes $(X_{\sigma^n \wedge t})_{t \in \Real_+}$ and $(X_{\tsigma^m \wedge t})_{t \in \Real_+}$ are submartingales of class D for all $n \in \Natural$ and $m \in \Natural$. (Note that we already know that $X = L - W$ is a local submartingale.) As a result,
 \begin{equation*}
 \begin{split}
  \lim_{n\rightarrow \infty} \expec[X_{\tau \wedge \sigma^n}]
  & = \lim_{n\rightarrow \infty} \expec\left[\lim_{m\rightarrow \infty} X_{\tau \wedge \sigma^n \wedge \tsigma^m} \right] \\
  & = \lim_{n\rightarrow \infty} \lim_{m\rightarrow \infty} \expec\left[ X_{\tau \wedge \sigma^n \wedge \tsigma^m} \right] \\
  & = \lim_{m \rightarrow \infty} \lim_{n \rightarrow \infty} \expec\left[ X_{\tau \wedge \sigma^n \wedge \tsigma^m} \right] \\
  & = \lim_{m \rightarrow \infty}  \expec\left[ \lim_{n \rightarrow \infty} X_{\tau \wedge \sigma^n \wedge \tsigma^m}  \right] \\
  & = \lim_{m\rightarrow \infty} \expec[X_{\tau \wedge \tsigma^m}].
 \end{split}
 \end{equation*}
Above, the limits in the third identity can be exchanged due to the fact that the double sequence $(\expec[X_{\tau \wedge \sigma^n \wedge \tsigma^m} ])_{n \in \Natural, \, m \in \Natural}$ is nondecreasing in both $n$ and $m$.

The fact that $\delta(\tau) \geq 0$ follows from the second identity in \eqref{eq: def delta}.
\end{proof}

\begin{rem}\label{remark: Z martingale}
Suppose that $\tau \in \Time$ is such that $\expec[Z_\tau] = Z_0 = 1$. In this case, $(Z_{\tau \wedge t})_{t \in \Real_+}$ is a uniformly integrable martingale and one can define a probability $\qprob \ll \prob$ on $(\Omega, \F_{\infty})$ via $\ud \qprob = Z_\tau \ud \prob$. Then, $\expec[X_{\tau}] = \expecq [\beta_\tau g(S_\tau)]$ (with conventions about the value of $\beta_\tau g(S_\tau)$ on $\set{\tau = \infty}$ similar to the ones discussed for $X$ previously). Let $\og :\Real_+ \mapsto \Real$ be defined as in the beginning of the proof of Lemma \ref{lemma:delta}. Since $\lim_{x \to \infty} (\og(x) / x) = 0$, there exists a nondecreasing function {\bf $\phi : \Real_+ \mapsto \Real_+ \cup  \{\infty\}$} such that $\phi(\og(x)) \leq x$ for all $x \in \Real_+$ and $\lim_{x \to \infty} (\phi(x) / x) = \infty$. Then,
\[
\sup_{\tau' \in \Time[0,\tau]} \expecq [\phi(\beta_{\tau'} \og(S_{\tau'}))] \leq \sup_{\tau' \in \Time[0,\tau]} \expecq [\phi( \og( \beta_{\tau'} S_{\tau'}))] \leq \sup_{\tau' \in \Time[0,\tau]} \expecq [\beta_{\tau'} S_{\tau'}] = S_0 < \infty.
\]
From de la Vall\'ee-Poussin's criterion, $\set{\beta_{\tau'} \og(S_{\tau'}) \such \tau' \in \Time[0,\tau]}$ is uniformly integrable with respect to $\qprob$. Then,
\begin{equation}\label{eq: delta simple}
\delta(\tau) = \limn \pare{\expec[L_{\tau \wedge \sigma^n}]  -   \expecq[ \beta_{\tau \wedge \sigma^n} \og(S_{\tau \wedge \sigma^n})]} - \pare{\expec[L_\tau] - \expecq[ \beta_{\tau} \og(S_{\tau})]} = L_0 - \expec[L_\tau]
\end{equation}
holds for any localizing sequence $(\sigma^n)_{n \in \Natural}$ of $L$. It follows that $\delta(\tau)$ is equal to the default of the local martingale $L$ at $\tau$.

As a consequence of the above observation, if $Z$ is a uniformly integrable martingale (and, in particular, if $Z \equiv 1$), the function $\delta$ is the same for all payoff functions $g$ (as long as $g$ satisfies the requirements specified in the Introduction, of course). This is no longer the case when $Z$ is not a uniformly integrable martingale, as we shall see later on in Example \ref{example:Z-local-L-local}.
\end{rem}

\subsection{A candidate for the smallest optimal stopping time}

We now aim at defining a stopping time $\tau^*$ that will be crucial in the solution of problem \eqref{eq: problem}. We need a preliminary result concerning a nice version of two processes $m$ and $M$ that have the following property: for all stopping times $\tau$, the random interval $[m_\tau, M_\tau]$ is the conditional support of $(S_t)_{t \in [\tau , \infty[}$ given $\F_\tau$.

\begin{lem}
Let $\uS_{\cdot} \dfn \inf_{t \in [\cdot, \infty[} S_t$ and $\oS_{\cdot} \dfn \sup_{t \in [\cdot, \infty[} S_t$. (The processes $\uS$ and $\oS$ are nonnegative, nondecreasing and nonincreasing respectively, \cadlag, and not adapted in general.) Then, there exists a nondecreasing  nonnegative \cadlag \ process $m$ and a nonincreasing nonnegative $[0, \infty]$-valued \cadlag \ adapted process $M$ such that (a) $m \leq \uS \leq \oS \leq M$, and (b) for all other processes $m'$ and $M'$ that share the exact same properties as $m$ and $M$ described previously, we have $m' \leq m \leq M \leq M'$.
\end{lem}

\begin{proof}
For all $t \in \Real_+$, let $\uC_t = \set{\xi \such \xi \text{ is } \F_t \text{-measurable and } \xi \leq \uS_t}$. Then, let $m^0_t$ denote the essential supremum of $\uC_t$, uniquely defined up to $\prob$-a.s. equality. Since $\uC_t$ is a directed set, $m^0_t \in \uC_t$. Furthermore, for all $t' \geq t$ we have  $m^0_t \in \uC_{t'}$; therefore, $\prob[m^0_t \leq m^0_{t'}] = 1$. Next, we will construct a c\`{a}dl\`{a}g modification of $m^0_t$.
Define
\[
m_t = \inf_{\mathbb{D} \ni t' > t} \sup_{\mathbb{D} \ni u \leq t'} m^0_u,
\]
where $\mathbb{D}$ is a countable and dense subset of $\Real_+$. As the filtration is right-continuous, we still have $m_t \in \F_t$ for $t \in \Real_+$. Also, $\prob[m_t \geq m^0_t] = 1$ holds for $t \in \Real_+$. The right-continuity of $S$ gives that $m_t \in \uC_t$ for $t \in \Real_+$; it follows that $\prob[m^0_t = m_t] = 1$ holds for all $t \in \Real_+$. By way of construction, $m$ is right-continuous and nondecreasing. We can define $M$ in a similar way, starting by setting $M^0_t$ be the essential infimum of $\oC_t = \set{\xi \such \xi \text{ is } \F_t \text{-measurable and } \oS_t \leq \xi}$ for $t \in \Real_+$, which might take the value $\infty$ with positive probability. It is quite straightforward to check that the two processes $m$ and $M$ have the properties in the statement of the lemma.
\end{proof}

Define $g' : \Real_+ \mapsto \Real_+$ via $g' (x) \dfn \ \downarrow \lim_{n \to \infty} n \pare{g(x + 1 / n) - g(x)}$ for $x \in \Real_+$, i.e., $g'$ is the right-derivative of $g$. The function $g'$ is right-continuous, nonnegative and nondecreasing. 
Let $K \dfn \sup \set{x \in \Real_+ \such g(x) = g'(0) x}$; in the case of a call option, $g'(0) = 0$ and $K$ corresponds to the strike price. We also define two functions $\ell: \Real_{++} \mapsto \Real_+$ and $r: \Real_{++} \mapsto \Real_{++} \cup \set{\infty}$ by setting
\[
\ell (x) \dfn \inf \set{y \in \Real_+ \such g'(y) = g'(x)}, \text{ as well as } r (x) \dfn \sup \set{y \in \Real_+ \such g'(y) = g'(x)}\]
for all $x \in \Real_{++}$. It is clear that $\ell$ and $r$ are nondecreasing functions, that $\ell(x) \leq x \leq r(x)$ for all $x \in \Real_{++}$, and that $g$ is affine on each interval $[\ell(x), \, r(x)]$ for $x \in \Real_{++}$.  It is also straightforward that there exists an at most countable collection of intervals $I_i = [\lambda_i, \rho_i] \subseteq \Real_+$, $i \in \Natural$, with $\lambda_i < \rho_i$, $]\lambda_i, \rho_i[ \cap ]\lambda_j, \rho_j[ = \emptyset$ for $\Natural \ni i \neq j \in \Natural$, and such that each $I_i$ is equal to $[\ell(x), r(x)]$ for some $x \in \Real_{++}$.


\smallskip

We are now ready to define the candidate $\tau^*$ for the minimal optimal stopping time for the problem \eqref{eq: problem}. To get an intuition for this, consider the case where $Z \equiv 1$. Sitting at some point in time, suppose that we know that $S$ will actually stay forever in $[0, K]$; in that case, and since $g(x) = g'(0) x$ holds for $x \in [0, K]$, $X$ will be a local martingale from that point onwards; therefore, we should stop. Further, suppose that we know that $S$ will never escape from an interval other than $[0, K]$ where $g$ is affine (which, of course, includes the case that $S$ remains a constant), as well as that $\beta$ will certainly not decrease further. We again have that the process $X$ will behave like a nonnegative local martingale from that point onwards, which implies that it is optimal to stop immediately. Therefore, we should certainly stop in either of the above cases.

We proceed now more rigorously in the construction of $\tau^*$. Define $\tau_K \dfn \inf \set{t \in \Real_+ \such M_t \leq K}$. Now,  if $\zeta$ denotes the \cadlag \ modification of the nonnegative supermartingale $(\beta_t - \expec[\beta_\infty \such \F_t])_{t \in \Real_+}$, define $\ttau := \inf\{t \in \Real_+ \such \zeta_t = 0\}$. On $\{\ttau < \infty\}$, we have $\zeta_{\ttau} = 0$, which is equivalent to $\beta_{\ttau} = \beta_{\infty}$ since $\beta$ is nonincreasing. For each $I_i = [\lambda_i, \rho_i]$, $i \in \Natural$, as described above, define the stopping time $\tau^i := \ttau \vee \inf\{t \in \Real_+ \such \lambda_i \leq m_t \leq M_t \leq \rho_i\}$. Finally, define the stopping time $\tau^{0} : = \inf\{t \in \Real_+ \such m_t = M_t\}$; observe that $\ttau \leq \tau^0$, since both $Z \beta S$ and $Z$ are local martingales. Finally, we define the following stopping time, which will be used in the main result:
\begin{equation} \label{eq: tau_star}
 \tau^* \dfn \tau_K \wedge \pare{ \bigwedge_{i \in \Natural \cup \{ 0 \}} \tau^i }.
\end{equation}
From the construction of $\tau^*$, it is clear that on $\set{\tau^* < \tau_K}$ we have $\beta_{\tau^*} =\beta_{\infty}$ and $\ell(S_{\tau^*}) \leq \inf_{t \in [\tau^*, \infty[}  S_{t} \leq \sup_{t \in [\tau^*, \infty[}  S_t \leq r(S_{\tau^*})$.

\subsection{The main result}

Here is our main result, whose proof is given in Section \ref{sec: proof}.

\begin{thm} \label{thm: main}
For the problem described in \eqref{eq: problem}, we have the following:
\begin{enumerate}
  \item The value of the problem is $v =  \expec[X_{\tau^*}] + \delta (\tau^*) = \expec[X_{\infty}] + \delta (\infty)$.
  \item A stopping time $\htau \in \Time$ is optimal  if and only if $\tau^* \leq \htau$, as well as $\delta(\htau) = 0$.
  \item Optimal stopping times exist if and only if $\delta (\tau^*) = 0$. In that case, $\tau^*$ is the smallest optimal stopping time, and the set of all optimal stopping times is $\set{\htau \in \Time \such \tau^* \leq \htau \text{ and } \delta(\htau) = 0}$.
\end{enumerate}
\end{thm}

\begin{rem}
 When the problem has a finite horizon $T\in \Time$, the value of \eqref{eq: problem} is
 \begin{equation}\label{eq:opt-value-finite-hor}
  v = \expec[X_{T}] + \delta (T).
 \end{equation}
When $\beta =1$, $\expec[Z_T]=1$, and $Z S$ is a continuous strict local martingale, \eqref{eq:opt-value-finite-hor} has been proved in Theorem~A.1 of \cite{Cox-Hobson} and in Proposition~2 in \cite{Madan-Yor}. Theorem~\ref{thm: main} (1) generalizes their result to a setting in which $\beta$ may not be 1, $S$ may have jumps, and $Z$ may not necessarily be a martingale.
\end{rem}

\begin{rem}

In proving Theorem~\ref{thm: main} (see Section \ref{sec: proof}), we take a bare-hands approach to the problem described in \eqref{eq: problem}, instead of using the well-developed theory of optimal stopping using Snell envelopes. The reason is the following: the most celebrated result from the optimal stopping theory (see \cite{ELKaroui} and Appendix~D of \cite{Karatzas-Shreve}) only gives a \emph{sufficient} condition that guarantees the existence of optimal stopping times. To use this result, we would have to assume that the process $X = \pare{Y_{t}g(S_t)}_{t \in \Real_+}$ is of class D. However, in general, this fails in problem \eqref{eq: problem}. For example when $Z \equiv 1$ and $g(x)=(x-K)_{+}$ for $x \in \Real_+$ (where $K \in \Real_{++}$), the process $\pare{\beta_{t}g(S_t)}_{t \in \Real_+}$ is of class D if and only if $\beta S$ is a martingale. This is insufficient for our purposes, since we are interested in cases where $\beta S$ is a strict local martingale.
\end{rem}

\section{Examples} \label{sec:examples}
Throughout this section we assume that the horizon of \eqref{eq: problem} is $T \in \Real_{++}$.
In \S\ref{sec:optimal-ex-time}, we give two examples and show that the smallest optimal exercise time may be equal to $T$ or strictly less than $T$. In \S\ref{sec:put-call-parity}, we discuss the implication of Theorem~\ref{thm: main} on the put-call parity. In \S \ref{sec:markovian}, we show that in a Markovian setting the American call price is one of infinitely many solutions of a Cauchy problem; we also indicate how one can uniquely identify the American call price using the put-call parity. In \S \ref{sec:frem}, we give an example to show that optimal exercise times may not exist and that the American call option price may always be strictly greater than the pay-off even at infinity, which further complicates the numerical pricing of American options using finite difference methods.

\subsection{Optimal exercise times}\label{sec:optimal-ex-time}
Here, we assume that the payoff $g$ is the call option payoff, i.e., $g(x) = (x-K)_+$ for $x \in \Real_+$, where $K\in \Real_{++}$ is the strike price. In this case, $\og(x) = x - g(x) = x \wedge K$ holds for $x \in \Real_+$.  Also, the interval $[\ell(S_t), r(S_t)]$ is either $[0,K]$ or $[K, \infty]$ for any $t\leq T$. In both examples below, we assume that the stochastic basis is rich enough to accommodate a process $Z$ which is the reciprocal of a 3-dimensional Bessel process starting from one; this is a classical example, due to \cite{Johnson-Helms}, of a strict local martingale.

\begin{exa}\label{example:Z-local-L-mart}

With $Z$ as above, define $\beta \equiv 1$ and $S = 1/ Z$; then $ L = Z\beta S \equiv 1$ is a martingale. It was shown in \cite{Delbaen-Schachermayer-Bessel} that arbitrage opportunities exist in a market with an asset $S$. Such an arbitrage opportunity was explicitly given in Example~4.6 of \cite{Karatzas-Kardaras}.

For $t < T$ and given $\F_t$ (and, therefore, $S_t$), $S$ crosses $K$ in $[t,T]$ with strictly positive probability, which can be shown using the explicit expression for its density given by
(1.06) of \cite{Borodin-Salminen} on page 429. Therefore $\tau^*=T$ is the only candidate for an optimal stopping time. We claim that $\delta(T)=0$, hence $T$ is the only optimal stopping time thanks to Theorem~\ref{thm: main} (3). The last claim is not hard to prove: since $L\equiv 1$, we can choose $\sigma^n=n$ for all $n\in \Natural$ as a localizing sequence for $L$; then, $\delta(T)  = \lim_{n\rightarrow \infty} \expec[X_{T\wedge n}] - \expec[X_{T}] = 0$, because for all $n \in \Natural$ with $n \geq T$ we have $\expec[X_{T\wedge n}] = \expec[X_{T}]$.
\end{exa}

The terminal time $T$ may not be the only candidate to exercise optimally. In the following example, it is optimal to exercise before $T$.

\begin{exa}\label{example:Z-local-L-local}
With $Z$ as above, define $K=2$, $S_t=\left\{\begin{array}{ll}1, & t< t_0;\\ 4, & t\geq t_0;\end{array}\right.$ and $\beta_t = \left\{\begin{array}{ll} 1, & t<t_0; \\ 1/4, & t\geq t_0;\end{array}\right.$ for all $t \in [0, T]$, where $t_0$ is a constant with $0 < t_0 < T$. Hence, both $Z$ and $L$ are strict local martingales. It is clear that $\tau^*=t_0$. Let $(\tsigma^n)_{n\in \Natural}$ be a sequence localizing $Z$ (and, since $L = Z$, localizing $L$ as well). We have from Lemma~\ref{lemma:delta} that
 \[
  \delta(t_0) = \lim_{n\rightarrow \infty} \expec\left[Y_{\tsigma^n} \left(S_{\tsigma^n} - K\right)_+ \, \indic_{\{t_0> \tsigma^n\}}\right] =0,
 \]
 because $S_{\tsigma^n}< K$ on $\{\tsigma^n < t_0\}$. Therefore, $t_0$ is the smallest optimal exercise time in view of Theorem~\ref{thm: main} (3).  The value of this problem is $v = \expec[Y_{t_0}\pare{S_{t_0}-K}_+] = \expec[Z_{t_0}]/2$. Moreover, with $\xi^n := \inf\set{t\geq t_0 \,:\, Z_t/Z_{t_0} \geq n}$ for $n\in \Natural$, we have $\expec[Z_{\xi^n}] = \expec[Z_{t_0}]$. It follows that any element of the sequence $\pare{\xi^n}_{n\in \Natural}$ is an optimal stopping time.

 It is also worth noticing that even though $t_0$ is an optimal exercise time, $\delta(t_0) < L_0 - \expec[L_{t_0}]$. (Compare this fact to \eqref{eq: delta simple} in Remark~\ref{remark: Z martingale}.) This observation follows from $L_0 - \expec[L_{t_0}] = 1-\expec[Z_{t_0}]>0$ and $\delta(t_0)=0$, which we have shown above.

Let us also use this example to show that the mapping $\delta$ depends on the form of the payoff function $g$.
For this purpose, let us pick another call option with strike price $\widehat{K}=1/2$. For this option,
$\delta(t_0) = (1/2) \, \lim_{n\rightarrow \infty} \expec\left[Z_{\tsigma^n} \indic_{\{t_0> \tsigma^n\}}\right]$, which is strictly positive. This is because
 \[\lim_{n\rightarrow \infty} \expec[Z_{\tsigma^n} \indic_{\{t_0>\tsigma^n\}}] = \lim_{n\rightarrow \infty} \expec[Z_{t_0 \wedge \tsigma^n}] - \lim_{n\rightarrow \infty} \expec[Z_{t_0} \indic_{\{t_0\leq \tsigma^n\}}] = 1- \expec[Z_{t_0}] >0,\] in which the second equality follows from the dominated convergence theorem.
 The dependence of $\delta$ on $g$ should be contrasted to \eqref{eq: delta simple} in Remark~\ref{remark: Z martingale}, which shows that $\delta$ does not depend on the form of $g$ when $Z$ is a uniformly integrable martingale.
 \qed
\end{exa}

\subsection{Put-call parity}\label{sec:put-call-parity}
If either the discounted stock price or the local martingale $Z$ is a strict local martingale, the put-call parity fails; see Theorem 3.4 (iii) in \cite{Cox-Hobson}, Section 3.2 in \cite{Heston-Loewenstein-Willard}, \cite{MR2359365}, and \cite{jps} when $Z$ is uniformly integrable and $Z \beta S$ is a strict local martingale. (In \cite{MR2359365}, it is shown that the put-call parity holds under
Merton's concept of no-dominance although it fails when only no-free-lunch with vanishing risk is assumed.)
In the case where $Z$ is a strict local martingale, see Remarks 9.1 and 9.3 in \cite{Fernholz-Karatzas-Kardaras} and Remark 10.1 in \cite{Fernholz-Karatzas-survey} .

We assume here that there exists a unique local martingale $Z$ that makes $Z \beta S$ a local martingale. We also assume that $Z$ is actually a martingale. One then can define a probability measure $\qprob \sim \prob$ on $(\Omega, \F_T)$ via $\ud \qprob = Z_T \ud\prob$. As a result, $\beta S$ is a local martingale under $\qprob$.
In this setting, we shall show below that the put-call parity still holds when the European call option is replaced by its American counterpart.

With an obvious extension of Theorem~\ref{thm: main} (1), and in par with Remark \ref{remark: Z martingale}, we obtain that the American option price process $A$ can be represented by
\begin{equation*}
 A_t = \expecq\left[\left(\beta_T / \beta_t\right) \, g(S_T) \,\big|\, \F_t\right] + \expecq\left[S_t - \left(\beta_T/ \beta_t\right)\, S_T \,\big|\, \F_t\right], \quad \text{for all } t\in [0,T].
\end{equation*}
Recalling that $\og(x) = x-g(x)$ for $x\in \Real_+$ from the proof of Lemma \ref{lemma:delta}, one has
\begin{equation*}
 A_t = S_t - \beta_t^{-1} \expecq\left[\beta_T \, \overline{g}(S_T) \, \big|\, \F_t\right].
\end{equation*}
 Now, if we denote $\overline{E}_t \dfn \beta_t^{-1} \expecq\left[\beta_T \, \overline{g}(S_T) \, \big|\, \F_t\right]$ for each $t \in [0,T]$, which the value at time $t$ of a European option with the payoff $\overline{g}(S_T)$ at maturity, we obtain
\begin{equation}\label{eq:put-call-parity}
 A+ \overline{E} = S.
\end{equation}
Therefore, we retrieve the put-call parity, but with an American option on the ``call" side instead of a European. Moreover, \eqref{eq:put-call-parity} helps us uniquely identify the American option price in the Markovian case in the next subsection.

\subsection{Characterizing the American option price in terms of PDEs} \label{sec:markovian}
Here, in addition to the all assumptions of \S \ref{sec:put-call-parity}, we further assume that the discounting process is $\beta = \exp\left(-\int_0^\cdot r (t) \, \ud t \right)$, where the interest rate process $r$ is nonnegative and deterministic, as well as that the dynamics of $S$ under the probability $\qprob$, are given by
\begin{equation}\label{eq:dyn-S}
 dS_t = r (t) S_t dt + \alpha(S_t) \, \ud B_t,
\end{equation}
where $B$ is a Brownian motion under $\qprob$. We assume that $\alpha(x) > 0$ for $x>0$, $\alpha(0)=0$, and that $\alpha$ is continuous and locally H\"{o}lder continuous with exponent 1/2; therefore, the stochastic differential equation \eqref{eq:dyn-S} has a unique strong nonnegative solution $S$. Additionally, the nonnegative volatility $\alpha$ satisfies $\int_1^{\infty} \pare{x / \alpha^2(x)} \ud x <\infty$, which is a necessary and sufficient condition for $\beta S$ to be strict local martingale under $\qprob$; see \cite{citeulike:64915}.

The price-process $A$ of an American option satisfies $A_t = a(S_t, t)$ for $t \in [0, T]$, where the value function $a : \Real_+ \times [0, T] \mapsto \Real_+$ is given by
\begin{equation}\label{eq:a-value}
a(x,t) \dfn x- \overline{e}(x,t),
\end{equation}
in which $\overline{e}(x,t) := \expecq \left[\left. \exp\left(- \int_t^T r(u)\, \ud u\right) \overline{g}(S_T) \right|S_t= x\right]$ for $(x, t) \in \Real_+ \times [0, T]$ is the value function of a European option with the payoff $\overline{g}$. Recall that $\lim_{x\rightarrow \infty} \og(x) / x =0$, i.e., $\overline{g}$ is a function of \textit{strictly sublinear growth}, according to Definition~4.2 in \cite{Ekstrom-Tysk}.

When the discounted stock price is a strict local martingale, the differential equation, satisfied by the European option price, usually has multiple solutions, see \cite{Heston-Loewenstein-Willard}. However,
when the terminal condition has strict sublinear growth the same PDEs have a unique solution in the class of functions with sublinear growth,  see Theorem~4.3 of \cite{Ekstrom-Tysk}. We will use this result to uniquely identify the price of American option in what follows. A simple modification\footnote{When $r\neq 0$, (17) in \cite{Ekstrom-Tysk} can be replaced by $v_t^{\epsilon} - \frac12 \alpha^2(x) v_{xx}^{\epsilon} - r x v_{x}^{\epsilon} + r v^{\epsilon} = \epsilon e^t (1+r+x) >0.$ The rest of the proof can be adapted in a straightforward manner to the nonzero interest rate case.} of Theorem~4.3 in \cite{Ekstrom-Tysk} to the nonzero interest rate case gives the following result.

\begin{prop}\label{prop:eq-e-bar}
 The value function $\overline{e}$ is the unique classical solution in the class of strictly sublinear growth functions of the following boundary value problem
 \begin{equation}\label{eq:e-bar}
 \begin{split}
  & \overline{e}_t + \frac12 \alpha^2(x)\, \overline{e}_{xx} + r(t)\, x \overline{e}_x - r(t)\, \overline{e} =0, \quad (x,t)\in \Real_+ \times [0,T),\\
  & \overline{e}(x,T) = \overline{g}(x),\\
  & \overline{e}(0,t) = 0, \quad 0\leq t < T.\\
 \end{split}
 \end{equation}
\end{prop}

\begin{rem}
Proposition~\ref{prop:eq-e-bar} uniquely characterizes the American option price. First one needs to find the solution of \eqref{eq:e-bar} with strictly sublinear growth in its first variable (which is unique). Subtracting this value from the stock price gives the American option price. The approximation method described by Theorem 2.2 of \cite{EST} can also be used to compute $\overline{e}$. (The idea is to approximate $\overline{e}$ by a sequence of functions that are unique solutions of Cauchy problems on bounded domains.)
\end{rem}

\begin{rem}
Thanks to fact that $\overline{e}$ is of strictly sublinear growth, \eqref{eq:a-value} gives $\lim_{x\rightarrow \infty} a(x,t) / x=1$ for $t\in [0,T]$, i.e., $a(x, t)$ is of linear growth in $x$. Moreover, Proposition \ref{prop:eq-e-bar} also implies that $a(x, t)$ is a classical solution of the following boundary value problem
\begin{equation}\label{eq:a}
\begin{split}
  & a_t + \frac12 \alpha^2(x)\, a_{xx} + r(t)\, x a_x - r(t)\, a =0, \quad (x,t)\in \Real_+ \times [0,T),\\
  & a(x,T) = g(x),\\
  & a(0,t) = 0, \quad 0\leq t < T.\\
\end{split}
\end{equation}
However, $a(x,t)$ is not the unique solution of \eqref{eq:a}. Indeed, consider the value of European option
\begin{equation*}
 e(x,t) \dfn \mathbb{E}_{\qprob}\left[\left. \exp\left(-\int_t^T r(u)\, \ud u\right) g(S_T) \right| S_t=x\right], \quad \text{ for all } (x,t)\in \Real_+ \times [0,T],
\end{equation*}
is another classical solution of \eqref{eq:a} (see Theorem~3.2 in \cite{Ekstrom-Tysk}). Moreover, since $\beta S$ is a strict local martingale under $\qprob$, we have
\begin{equation*}
 a(x,t) - e(x,t) = x- \mathbb{E}_{\qprob}\left[\left. \exp\left(-\int_t^T r(u) \, \ud u\right) S_T \right| S_t = x\right] > 0, \quad \text{ for all } (x,t)\in \Real_+ \times [0,T],
\end{equation*}
i.e., the American option value dominates its European counterpart, which is  the smallest nonnegative supersolution of \eqref{eq:a}. (See the comment after Theorem~4.3 in \cite{Ekstrom-Tysk}.)

It is also worth-noting that $a-e$ satisfies \eqref{eq:a} with zero as the terminal condition.
Thanks to this observation, we can construct infinitely many solutions to \eqref{eq:a} greater than the price of the American option. That is to say, the American option price is neither the smallest nor the largest solution of \eqref{eq:a}. Similarly, we can also fill the gap between the European and American options with infinitely many solutions of \eqref{eq:a}. In fact, this Cauchy problem has a unique solution among functions of linear growth if and only if $\beta S$ is a martingale under $\qprob$; see \cite{bx09}.
\end{rem}

\subsection{A further remark}\label{sec:frem}

In addition to all assumptions of \S~\ref{sec:markovian}, we assume that $\beta \equiv 1$, $g(x)=(x-K)_+$ for $x \in \Real_+$ where $K \in \Real_{++}$, and $S$ is the reciprocal of a 3-dimensional Bessel process. In this case, an argument similar to the one in Example~\ref{example:Z-local-L-mart} gives $\tau^*=T$. Moreover, it follows from \eqref{eq: delta simple} that $\delta(T) >0$. Therefore, thanks to Theorem~\ref{thm: main} (3), there is no optimal stopping time solving \eqref{eq: problem} in this setting.

We will demonstrate that the American call price is strictly larger than the payoff even when the stock price variable tends to infinity.

Recall that $a(x,t) - e(x,t) = x- \expecq\left[S_T \, \big|\, S_t=x\right]$. Here
\begin{equation}\label{eq:a-e-example}
 x- \expecq\left[S_T \, \big|\, S_t=x\right] = 2x \, \Phi\left(- \frac{1}{x\sqrt{T-t}}\right),
\end{equation}
where $\Phi(\cdot) = \frac{1}{\sqrt{2\pi}} \int_{-\infty}^{\cdot} e^{- x^2/2} dx$. (See Section~2.2.2 of \cite{Cox-Hobson}.) Meanwhile, Example~3.5 of \cite{Cox-Hobson} gives
\begin{equation}\label{eq:lim-e-example}
 \lim_{x\uparrow \infty} e(x,t) = \frac{2}{\sqrt{2\pi (T-t)}} - K \left[2\Phi\left(\frac{1}{K\sqrt{T-t}}\right) -1\right].
\end{equation}
Combining \eqref{eq:a-e-example} and \eqref{eq:lim-e-example}, we obtain
\begin{equation*}
\begin{split}
 \lim_{x\uparrow \infty} \left[a(x,t) - (x-K)_+\right] & = \lim_{x\uparrow \infty} e(x,t) - \lim_{x\uparrow \infty} \expecq\left[S_T \, \big|\, S_t=x\right] + K\\
 &= 2K \left[1-\Phi\left(\frac{1}{K\sqrt{T-t}}\right)\right] >0, \quad \text{ for all } t\in [0,T).
\end{split}
\end{equation*}

\section{Proof of Theorem \ref{thm: main}} \label{sec: proof}

\subsection{Proof of statement (1)}

We first show that $\expec[X_{\tau}] \leq \expec[X_{\tau^* \wedge \tau}]$ holds for all $\tau \in \Time$. On $\set{\tau^* = \tau_K} \cap \{\tau^*<\infty\}$,
we have $g(S_{\tau^* \vee t}) = g'(0) S_{\tau^* \vee t}$ for all $t \in \Real_+$. Therefore, $X_{\tau^* \vee t} = g'(0) Z_{\tau^* \vee t} \beta_{\tau^* \vee t} S_{\tau^* \vee t} = g'(0) L_{\tau^* \vee t}$
for all $t \in \Real_+$ on $\set{\tau^* = \tau_K} \cap \{\tau^*<\infty\}$. As $L$ is a nonnegative local martingale, therefore a supermartingale, $\expec[X_\tau \such \F_{\tau^*}] \leq X_{\tau^*}$
on $\set{\tau^* = \tau_K < \tau}$. Now, on $\set{\tau_K > \tau^*}$, by definition of $\tau^*$, we have  $g(S_{\tau^* \vee t }) = g(S_{\tau^*}) + g'(S_{\tau^*}) (S_{\tau^* \vee t} - S_{\tau^*})$ and $\beta_{\tau^* \vee t} = \beta_{\tau^*}$ for all $t \in \Real_+$. As $\set{\tau^* < \tau_K }\subseteq \set{X_{\tau^*} > 0}$, it follows that on the event $\set{\tau^* < \tau_K}$ we have
\begin{equation} \label{eq: local mart after tau*}
\frac{X_{\tau^* \vee \cdot}}{X_{\tau^*}} =  \pare{ \frac{\beta_{\tau^*} g(S_{\tau^*}) - \beta_{\tau^*} S_{\tau^*} g'(S_{\tau^*})}{X_{\tau^*}} } Z_{\tau^* \vee \cdot}    +  \pare{ \frac{g'(S_{\tau^*})}{X_{\tau^*}} } L_{\tau^* \vee \cdot},
\end{equation}
%
Since both $Z$ and $L$ are local martingales and $X$ is nonnegative, it is straightforward that the inequality $\expec[X_\tau \such \F_{\tau^*}] \leq X_{\tau^*}$ holds on $\set{\tau^* < \tau} \cap \set{\tau^* < \tau_K}$. Since $\expec[X_\tau \such \F_{\tau^*}] \leq X_{\tau^*}$ also holds on $\set{\tau^* < \tau} \cap \set{\tau^* = \tau_K}$, we obtain $\expec[X_\tau \such \F_{\tau^*}] \leq X_{\tau^*}$ on $\set{\tau^* < \tau}$, and $\expec[X_{\tau}] \leq \expec[X_{\tau^*\wedge \tau}]$ follows.

Let $(\sigma^n)_{n \in \Natural}$ be a localizing sequence for $L$. In the proof of Lemma \ref{lemma:delta} it was shown that  $(X_{\sigma^n \wedge t})_{t \in \Real_+}$ is a submartingale of class D. Therefore, $\sup_{\tau \in \Time{[0, \sigma^n]}} \expec[X_{\tau}] = \expec[X_{\sigma^n}]$. From the discussion of the previous paragraph, {\bf $\sup_{\tau \in \Time{[0, \sigma^n]}} \expec[X_{\tau}] = \expec[X_{\sigma^n}] \leq \expec [X_{\sigma^n \wedge \tau^*}] = \sup_{\tau \in \Time{[0, \sigma^n \wedge \tau^*]}} \expec [X_{\tau}]$}, where the last equality follows again from the fact that $(X_{\sigma^n \wedge \tau^* \wedge t})_{t \in \Real_+}$ is a submartingale of class D. The other inequality $\sup_{\tau \in \Time{[0, \sigma^n \wedge \tau^*]}} \expec[X_{\tau}] \leq \sup_{\tau \in \Time{[0, \sigma^n]}} \expec [X_{\tau}]$ trivially holds. Furthermore, $\sup_{\tau \in \Time} \expec[X_{\tau}] = \, \uparrow \limn \sup_{\tau \in \Time{[0, \sigma^n ]}} \expec [X_{\tau}]$ is easily seen to hold by a use of Fatou's lemma. Putting everything together, we have
\begin{equation} \label{eq: value, one}
v = \sup_{\tau \in \Time} \expec[X_{\tau}] = \, \uparrow \limn \sup_{\tau \in \Time{[0, \sigma^n ]}} \expec [X_{\tau}] = \, \uparrow \limn \expec [X_{\sigma^n \wedge \tau^*}] = \expec [X_{\tau^*}] + \delta(\tau^*),
\end{equation}
where the last equality holds by the definition of $\delta$.

The equality $v = \expec[X_{\infty}] + \delta(\infty)$ is proved similarly, replacing $\tau^*$ by $\infty$ in \eqref{eq: value, one}.

\subsection{Proof of statement (2)}

We start with the following helpful result.

\begin{prop} \label{prop: UI sup-mart}
For some $\tau \in \Time$, the following statements are equivalent:
\begin{enumerate}
	\item $\delta(\tau) = 0$.
	\item $\delta(\tau') = 0$ for all $\tau' \in \Time{[0 ,\tau]}$.
	\item The process $(X_{\tau \wedge t})_{t \in \Real_+}$ is a  submartingale of class D.
\end{enumerate}
\end{prop}

\begin{proof}
As the implications $(3) \Rightarrow (2)$ and $(2) \Rightarrow (1)$ are straightforward, we focus in the sequel on proving the implication $(1) \Rightarrow (3)$.

Since $X=L-W$, in which $L$ is a local martingale and $W$ is a supermartingale, $(X_{\tau \wedge t})_{t \in \Real_+}$ is a local submartingale. Therefore, we only need to show that $\set{X_{\tau'} \such \tau' \in \Time{[0, \tau]}}$ is uniformly integrable. Let $\tau' \in \Time{[0, \tau]}$ and $A \in \F_{\tau'}$. Then, with $\tau'' \dfn \tau' \indic_A + \tau \indic_{\Omega \setminus A}$, we have $\tau'' \in \Time{[0, \tau]}$ as well.
Let $(\sigma^n)_{n \in \Natural}$ be a localizing sequence for $L$. Once again we use the fact (proved in Lemma \ref{lemma:delta}) that $(X_{\sigma^n \wedge t})_{t \in \Real_+}$ is a submartingale of class D to obtain
\[
\expec[X_{\tau''}] \leq \liminf_{n \to \infty} \expec[X_{\tau'' \wedge \sigma^n}] \leq \liminf_{n \to \infty} \expec[X_{\tau \wedge \sigma^n}] = \expec[X_{\tau}],
\]
the last equality following from the fact that $\delta(\tau) = 0$. The inequality above is equivalent to $\expec[X_{\tau'} \indic_A] \leq \expec[X_{\tau} \indic_A]$. Since $A \in \F_{\tau'}$ was arbitrary, we get $X_{\tau'} \leq \expec[X_{\tau} \such \F_{\tau'}]$. Since $X \geq 0$ and $\expec[X_\tau] < \infty$, this readily shows that $\set{X_{\tau'} \such \tau' \in \Time{[0, \tau]}}$ is uniformly integrable.
\end{proof}

We now proceed with the proof of statement (2) of Theorem \ref{thm: main}. A combination of the results of Lemma \ref{lem: 1 -> 2 (i)} and Lemma \ref{lem: 1 -> 2 (ii)} below show that if a stopping time $\htau \in \Time$ is optimal then $\delta(\htau) = 0$, as well as $\tau^* \leq \htau $. In Lemma \ref{lem: 2 -> 1}, the converse of the previous statement is proved.

\begin{lem} \label{lem: 1 -> 2 (i)}
If $\htau$ is any optimal stopping time, then $\delta(\htau) = 0$.
\end{lem}

\begin{proof}
Let $(\sigma^n)_{n \in \Natural}$ be a localizing sequence for $L$. To begin with, $\expec[X_{\htau}] \leq \liminfn \expec[X_{\htau \wedge \sigma^n}]$ holds in view of Fatou's lemma. On the other hand, the optimality of $\htau$ gives $\limsupn \expec[X_{\htau \wedge \sigma^n}] \leq \expec[X_{\htau}]$. Therefore, $\expec[X_{\htau}] = \limn \expec[X_{\htau \wedge \sigma^n}]$. By the definition of $\delta$, the latter equality is equivalent to $\delta(\htau) = 0$.
\end{proof}

\begin{lem} \label{lem: 1 -> 2 (ii)}
If  $\tau \in \Time$ is such that $\prob[\tau < \tau^* ] > 0$, then there exists $\tau' \in \Time$ with $\expec[X_{\tau}] < \expec[X_{\tau'}]$.
\end{lem}

\begin{proof}

We have shown in the proof of statement (1) that $\expec [X_{\tau}] \leq \expec [X_{\tau^* \wedge \tau}]$ for all $\tau \in \Time$. Therefore, it is enough to prove that if  $\tau \in \Time[0,\tau^*]$ is such that $\prob[\tau < \tau^* ] > 0$, then there exists $\tau' \in \Time$ with $\expec [X_{\tau}] < \expec[X_{\tau'}]$. For the rest of the proof of Lemma \ref{lem: 1 -> 2 (ii)}, we fix $\tau \in \Time[0,\tau^*]$ with $\prob[\tau < \tau^* ] > 0$.

Let us first introduce some notation. Define the local martingale $L^{(\tau)}_{\cdot} := (L_{\tau \vee \cdot} / L_{\tau}) \indic_{\set{\tau < \infty}} + \indic_{\set{\tau =\infty}}$. Observe that $L^{(\tau)}$ is really the local martingale $L$ \emph{started} at $\tau$: we have $L^{(\tau)}_t = 1$ on $\set{t<\tau}$ and $L^{(\tau)}_t = L_t / L_{\tau}$ on $\set{\tau \leq t}$. 
Let $(\eta^n)_{n\in \Natural}$ be a sequence of stopping times defined by $\eta^1 = \tau$ and $\eta^n := \inf \big\{ t \geq \tau \such L^{(\tau)}_t \geq n \big\}$ for $n>1$. One obtains that $\eta^n \geq \tau$, $\uparrow \limn \eta^n = \infty$, as well as the fact that $\big\{ L^{(\tau)}_{\eta^n \wedge \sigma} \such \sigma \in \Time \big\}$ is a uniformly integrable family, which means that $(L^{\tau}_{\eta^{n}\wedge t })_{t \in \Real_+}$ is a uniformly integrable martingale. (To see the last fact, observe that $\expec[\sup_{\sigma \in \Time} L^{(\tau)}_{\eta^n \wedge \sigma}] \leq n + \expec[L^{(\tau)}_{\eta^n}] \leq n+1$, where the last inequality follows from the fact that $L^{(\tau)}$ is a nonnegative local martingale with $L^{(\tau)}_0 = 1$.) In particular, $\expec[L_{\sigma} \such \F_\tau] = L_{\tau}$ holds for all $\sigma \in \Time [\tau, \eta^n]$, $n \in \Natural$.

\smallskip

Below, we shall find $\tau' \in \Time$ such that $\expec[X_{\tau}] < \expec[X_{\tau'}]$ in two distinct cases:

\noindent \underline{Case 1}: Assume that $\prob \bra{\{\tau<\tau^*\} \cap \{S_{\tau} \leq K\}}>0$. In view of the definition of $\tau^*$ (and of $\tau_K$), for small enough $\epsilon>0$, the stopping time
\[
 \tau'' = \inf\{t\geq \tau \such S_t > K+\epsilon\}
\]
is such that $\tau \leq \tau''$ and $\prob \bra{\set{S_{\tau} \leq K} \cap \set{\tau < \tau''<\infty}}>0$. Letting $\tau' = \indic_{\{S_{\tau}> K\}} \tau + \indic_{\{S_{\tau}\leq K\}} \tau''\wedge \eta^n$ for large enough $n\in \Natural$, we have $\tau \leq \tau'$, $\expec[L_{\tau'} \such \F_\tau] = L_{\tau}$, as well as $\prob \bra{\set{S_{\tau'} > K} \cap \set{\tau < \tau'<\infty}}>0$. Observe that on $\set{\tau < \tau'} \subseteq \set{S_\tau \leq K}$, we have $X_{\tau} = Y_\tau g(S_\tau) = Y_\tau g'(0) S_\tau = g'(0) L_\tau$. On the other hand, $X_{\tau'} = Y_{\tau'} g(S_{\tau'}) \geq Y_{\tau'} g'(0) S_{\tau'} = g'(0) L_{\tau'}$ with $X_{\tau'} > g'(0) L_{\tau'}$ holding on $\set{S_{\tau'} >K} \cap \set{\tau < \tau' < \infty}$. Therefore,
$\expec[X_{\tau'} \indic_{\set{\tau<\tau'}}] > g'(0)\expec[L_{\tau'} \indic_{\set{\tau<\tau'}}] = g'(0) \expec[L_{\tau} \indic_{\set{\tau<\tau'}}] = \expec[X_{\tau} \indic_{\set{\tau<\tau'}}]$. Since $\tau \leq \tau'$, we obtain $\expec[X_{\tau'}] > \expec[ X_{\tau}]$.

%

\noindent \underline{Case 2}: Assume now that $\prob\bra{\{\tau<\tau^*\} \cap \{S_{\tau} \leq K\}}= 0$. Since $S_{\tau}>K$ on $\set{\tau<\tau^*}$ and using the definition of $\tau^*$ in \eqref{eq: tau_star}, we can find small enough $\epsilon>0$ such that the stopping time
\[
 \tau'' = \inf\set{t \geq \tau \,:\, \beta_t < \beta_{\tau} -\epsilon, \text{ or } S_t < \ell(S_\tau)-\epsilon, \text{ or } S_t > r(S_{\tau})+ \epsilon}
\]
satisfies $\tau''\geq \tau$ and $\prob\bra{(\set{\beta_{\tau} > \beta_{\tau''}} \cup \set{S_{\tau''}< \ell(S_\tau)} \cup \set{r(S_{\tau}) < S_{\tau''}}) \cap \set{\tau''<\infty}} >0$. Then letting $\tau'= \tau''\wedge \eta^n$ for large enough $n\in \Natural$, we have $\tau \leq \tau'$, $\expec[L_{\tau'} \such \F_\tau] = L_{\tau}$, as well as $\prob\bra{(\set{\beta_{\tau} > \beta_{\tau'}} \cup \set{S_{\tau'}< \ell(S_\tau)} \cup \set{r(S_{\tau}) < S_{\tau'}}) \cap \set{\tau'<\infty}} >0$. 

The convexity of $g$ gives
\begin{align} \label{eq:th-t-con}
 Z_{\tau'} \beta_{\tau'} g(S_{\tau'}) & \ \geq \ Z_{\tau'} \beta_{\tau'} g(S_{\tau}) + Z_{\tau'}\beta_{\tau'} g'(S_{\tau}) (S_{\tau'} - S_{\tau}) \\
\nonumber & \ = \ Z_{\tau'} \beta_{\tau'} \pare{g(S_{\tau}) - g'(S_{\tau})S_{\tau}} + Z_{\tau'} \beta_{\tau'} g'(S_{\tau}) S_{\tau'} \\
\nonumber  & \ \geq \ Z_{\tau'} \beta_{\tau} \pare{g(S_{\tau}) - g'(S_{\tau})S_{\tau}} + Z_{\tau'}\beta_{\tau'} g'(S_{\tau}) S_{\tau'}.
\end{align}
In \ref{eq:th-t-con} above, the first inequality is strict on $(\set{S_{\tau'} < \ell(S_{\tau})} \cup \set{r(S_{\tau}) < S_{\tau'}})\cap \set{\tau'<\infty}$. (Note that $Z_{\tau'} \beta_{\tau'}>0$ on $\set{\tau'<\infty}$.) On the other hand, the second inequality in \eqref{eq:th-t-con} is strict on $\set{\beta_{\tau'} < \beta_{\tau}} \cap \set{\tau'<\infty}$, since $Z_{\tau'}>0$ on $\set{\tau'<\infty}$ and $g(S_{\tau}) < g'(S_{\tau}) S_{\tau}$ when $S_{\tau}>K$. Given all the previous observations, we take expectations in \eqref{eq:th-t-con} to obtain 
\[
\begin{split}
\expec[X_{\tau'}] & \ > \ \expec[Z_{\tau'} \beta_{\tau} \pare{g(S_{\tau}) - g'(S_{\tau})S_{\tau}}] + \expec[Z_{\tau'}\beta_{\tau'} g'(S_{\tau}) S_{\tau'}]\\
& \ \geq \ \expec[Z_{\tau} \beta_{\tau} \pare{g(S_{\tau}) - g'(S_{\tau})S_{\tau}}] + \expec[Z_{\tau'}\beta_{\tau'} g'(S_{\tau}) S_{\tau'}]\\
& \ = \ \expec[X_{\tau}] + \expec[g'(S_{\tau}) \pare{L_{\tau'} - L_{\tau}}] \ = \ \expec[X_{\tau}],
\end{split}
\]
in which the second inequality follows since $Z$ is a supermartingale and $g(S_{\tau}) - g'(S_{\tau})S_{\tau}\leq 0$..
\end{proof}

\begin{lem} \label{lem: 2 -> 1}
If $\htau \in \Time$ is such that $\tau^* \leq \htau$ and $\delta(\htau) = 0$, then $\htau$ is optimal.
\end{lem}

\begin{proof}

Fix $\tau \in \Time$ and $\htau \in \Time$ with $\tau^* \leq \htau$ and $\delta(\htau) = 0$. We shall show that $\expec[X_{\tau}] \leq \expec[X_{\htau}]$.

Since $\delta(\htau) = 0$, $(X_{{\htau} \wedge t})_{t \in \Real_+}$ is a submartingale of class D in view of Proposition \ref{prop: UI sup-mart}. Then, $X_\tau \indic_{\set{\tau \leq {\htau}}}  \leq \expec [X_{\htau} \such \F_\tau] \indic_{\set{\tau \leq {\htau}}} = \expec [X_{{\htau}}  \indic_{\set{\tau \leq {\htau}}} \such \F_\tau]$ follows immediately. Taking expectations in the previous inequality, we obtain
\begin{equation} \label{ineq: 1}
\expec[X_\tau  \indic_{\set{\tau \leq {\htau}}} ] \leq \expec [X_{{\htau}}  \indic_{\set{\tau \leq {\htau}}} ]
\end{equation}

On $\set{\htau < \tau}$, we can use the same idea as in \eqref{eq: local mart after tau*} and the discussion following (with $\htau$ replacing $\tau^*$ and using the fact that $\tau^* \leq \htau$) to obtain $\expec[X_\tau  \such \F_{\htau}] \indic_{\set{{\htau} < \tau}} \leq X_{\htau} \indic_{\set{{\htau} < \tau}}$; in particular,
\begin{equation} \label{ineq: 2}
\expec[X_\tau \indic_{\set{{\htau} < \tau}}] \leq \expec[X_{\htau} \indic_{\set{{\htau} < \tau}}].
\end{equation}
Combining the inequalities \eqref{ineq: 1} and \eqref{ineq: 2}, we obtain $ \expec[X_\tau ] \leq \expec [X_{{\htau}} ]$.
\end{proof}

\subsection{Proof of statement (3)}

If an optimal stopping time $\htau \in \Time$ exists, then $\tau^* \leq \htau$ is true by statement (2) of Theorem \ref{thm: main}; then, $\delta (\tau^*) = 0$ follows from $\delta (\htau) = 0$, in view of Proposition \ref{prop: UI sup-mart}. Conversely, if $\delta (\tau^*) = 0$, then $\expec[X_{\tau^*}] = \expec[X_{\tau^*}] + \delta (\tau^*) = v$ and, therefore, $\tau^*$ is optimal. The fact that $\tau^*$ is the smallest optimal stopping time, if optimal stopping times exist, as well as the respresentation of the set of all possible optimal stopping times, also follows from statement (2) of Theorem \ref{thm: main}.

\bibliographystyle{siam}
\bibliography{biblio}
\end{document}